%% file: OPPIMRCfinal.tex
\documentclass[conference]{IEEEtran} 

\usepackage{epsfig} 
\usepackage{graphicx}
\usepackage{amssymb}
\usepackage{amsfonts}
\usepackage{latexsym}
\usepackage[cmex10]{amsmath}
\usepackage[usenames]{color}
\usepackage{array}
\usepackage{cite} 
\usepackage{import}  
\usepackage[squaren]{SIunits}
\usepackage{grffile}
\usepackage{multirow}
\usepackage{booktabs} 
\usepackage{epstopdf}
\usepackage{import}
\usepackage{color}
\usepackage[letterpaper, left=1.05in, right=1.05in, bottom=1in, top=0.75in]{geometry}

\def\mathlette#1#2{{\mathchoice{\mbox{#1$\displaystyle #2$}}%
                               {\mbox{#1$\textstyle #2$}}%
                               {\mbox{#1$\scriptstyle #2$}}%
                               {\mbox{#1$\scriptscriptstyle #2$}}}}
\renewcommand{\Vec}[1]{\mathlette{\boldmath}{#1}}

\newcommand{\be}{\begin{equation}}
\newcommand{\ee}{\end{equation}}
\newcommand{\ba}{\begin{array}}
\newcommand{\ea}{\end{array}}
\newcommand{\bdm}{\begin{displaymath}}
\newcommand{\edm}{\end{displaymath}}
\newcommand{\bea}{\begin{eqnarray}}
\newcommand{\eea}{\end{eqnarray}}
\newcommand{\bean}{\begin{eqnarray*}}
\newcommand{\eean}{\end{eqnarray*}}
\newcommand{\me}{\text{e}}

\newcommand{\mj}{\text{j}}

\def\argmin{\mathop{\text{argmin}}}

\def\diag{\text{diag}}

\def\dif{\text{d}\,}
\def\E{\mathop{{}\mathbb{E}}}
\def\vects{\text{vec}}

\def\nTC{\ensuremath{T_\text{C}}} 
\def\nTS{\ensuremath{T_\text{S}}} 

\def\oH{\ensuremath{^\text{H}}} 
\def\oT{\ensuremath{^\text{T}}} 

\def\nfC{\ensuremath{f_\text{C}}}        

\def\nnuD{\ensuremath{\nu_\text{D}}}     
\def\nthetaP{\ensuremath{\theta_\text{P}}}  

\def\nEb{\ensuremath{E_\text{b}}}        

\IEEEoverridecommandlockouts

\title{Iterative Detection for Orthogonal Precoding\\
in Doubly Selective Channels}

\author{
\IEEEauthorblockN{Thomas Zemen, Markus Hofer, David L\"oschenbrand and Christoph Pacher}
\IEEEauthorblockA{\textit{Security \& Communication Technologies, Center for Digital Safety \& Security}}
\textit{AIT Austrian Institute of Technology}\\
Vienna, Austria\\
\{thomas.zemen, markus.hofer, david.loeschenbrand, christoph.pacher\}@ait.ac.at}

\begin{document}

\maketitle

\begin{abstract}
Ultra-reliable wireless communication links require the utilization of all diversity sources of a wireless communication channel. Hadani et al. propose a two dimensional discrete symplectic Fourier transform (DSFT) as orthogonal pre-coder for a time-frequency modulation scheme. In this paper we explore \emph{general} orthogonal precoding (OP) and its performance in time- and frequency-selective channels. We show that iterative parallel interference cancellation (PIC) and iterative channel estimation methods can be used for the detection of OP. A scalar signal model for OP transmission is obtained by PIC. Based on this signal model, we can prove that all constant modulus sequences, e.g. the DSFT basis functions or Walsh-Hadamard sequences, lead to the same performance for OP. We validate our receiver structure by numerical link level simulations of a vehicle-to-vehicle communication link with a relative velocity of $0\ldots200\,\text{km/h}$. We demonstrate that OP achieves a gain of about $4.8\,\text{dB}$ if compared to orthogonal frequency division multiplexing at a bit error rate of $10^{-4}$. Our performance results for coded OP are the best results for a fully documented receiver architecture, published so far.
\end{abstract}

\section{Introduction}
\label{sec:intro}
Reliable wireless communication links in time- and frequency-selective channels are a key requirement for future 5G system. Connected autonomous vehicles and industry 4.0 production environments are exemplary (5G) applications that require an ultra-reliable low-latency communication (URLLC) link \cite{Shafi17}. 

Linear precoding (or spreading) of data symbols in combination with orthogonal frequency division multiplexing (OFDM) is proposed by multiple authors in the context of multi-carrier (MC) code-division multiple access (CDMA) \cite{Linnartz01, Svensson04, Persson02} to obtain a diversity gain in the time- and/or frequency-domain. 

Orthogonal time frequency signaling (OTFS) \cite{Hadani17} focuses on the utilization of time- and frequency diversity. OTFS uses a 2D discrete symplectic Fourier transform (DSFT) to precode data symbols at the transmitter (TX) side. Hence, the information of each data-symbol is linearly spread on all available grid points of a data frame. After the precoding operation, the data frame is transmitted with a time-frequency modulation technique. At both, the TX and RX side, optional window functions can be applied.First performance results for OTFS and a comparison with OFDM are shown in \cite{Hadani17}. Publications \cite{Hadani17, Hadani18} present \emph{no details of their RX algorithm}, only performance results are shown. 

A message passing algorithm is applied for OTFS detection, exploiting sparse channels, in \cite{Raviteja17}. Only uncoded performance results are shown in \cite{Raviteja17}. A signal model for OTFS using OFDM is presented in \cite{RezazadehReyhani17}. A simplified OTFS modem structure for the case of rectangular transmit and receive windowing is shown in  \cite{Farhang17}. No performance results are given in both publications \cite{RezazadehReyhani17,Farhang17}. An equalization method utilizing the delay-Doppler representation of the DSFT is proposed in \cite{Li17}.

All recent work mentioned above, focuses on orthogonal precoding (OP) with the DSFT. It is currently unknown which OP basis provides the best performance in doubly selective channels. For this research area, a detection algorithm for general OP schemes in doubly selective channels is missing, and therefore, is the focus of this paper. 

\subsection*{Contributions of the Paper:} 
\begin{itemize}
\item We generalize the signal model presented in \cite{RezazadehReyhani17}, to general orthonormal basis functions. Parallel interference cancellation (PIC) \cite{Zemen06c} and iterative channel estimation methods \cite{Zemen12a} are applied for OP. We present an iterative RX for OP using \emph{coded transmission} for the first time. This fully documented receiver structure enables independent research result verification. Our RX provides the best performance results for coded OP using a fully documented RX algorithm, up to our best knowledge.

\item With the PIC structure of our RX, we obtain a scalar signal model for each grid point of the OP transmission. We prove that any complete set of basis functions with constant modulus will achieve the same performance for OP. Furthermore, we quantify analytically the channel hardening effect of OP for doubly-selective channels.

\item We validate our theoretical results by numerical link level simulations for a vehicle-to-vehicle communication link, using the IEEE 802.11p physical layer with an improved pilot pattern.
\end{itemize} 

\subsection*{Notation:}
We denote a scalar by $a$, a column vector by $\Vec{a}$ and its $i$-th element with $a_i$. Similarly, we denote a matrix by $\Vec{A}$ and its $(i,\ell)$-th element by $a_{i,\ell}$. The transpose of $\Vec{A}$ is given by $\Vec{A}\oT$ and its conjugate transpose by $\Vec{A}\oH$. A diagonal matrix with elements $a_i$ is written as $\diag(\Vec{a})$ and the $Q\times Q$ identity matrix as $\Vec{I}_Q$, respectivelly. The absolute value of $a$ is denoted by $\left|a\right|$ and its complex conjugate by $a^*$. The number of elements of set $\mathcal{I}$ is denoted by $|\mathcal{I}|$. We denote the set of all complex numbers by $\mathbb{C}$. The all one (zero) column vector with $Q$ elements is denoted by $\Vec{1}_Q$ ($\Vec{0}_Q$). We identify the 2D sequence $(a_{i,\ell})\in\mathbb{C}^{N \times M}$ for $i\in\{0,\ldots, N-1\}, \ell\in\{0,\ldots, M-1\}$ with the matrix $\Vec{A}\in\mathbb{C}^{N \times M}$, i.e., $\Vec{A}=(a_{i,\ell})$. Furthermore, we define the notation $\Vec{a}=\vects(\Vec{A})=\vects\big((a_{i,\ell})\big)\in\mathbb{C}^{MN \times 1}$, where $\vects(\Vec{A})$ denotes the vectorized version of matrix $\Vec{A}$, formed by stacking the columns of $\Vec{A}$ into a single column vector.

\section{Signal Model for General OP}
\label{se:SignalModel}

A system model for OTFS using OFDM as time-frequency modulation is introduced in \cite[(32)]{RezazadehReyhani17}. We generalize this model for arbitrary OP sequences in doubly selective channels. For now, we exclude channel estimation to simplify the exposition for Sections \ref{se:SignalModel}-\ref{sec:EffectiveChannel}. Later in Section \ref{sec:ChannelEstimation}, data and pilot multiplexing as well as iterative channel estimation for OP are introduced.

\subsection{General Orthogonal Precoding}
We precode data symbols $b_{p,n}\in\mathcal{A}$, $p\in\{0,\ldots, N-1\}$, $n\in\{0,\ldots, M-1\}$ on a general transform domain grid, with a general complete set of 2D orthonormal basis functions:
\be
d_{q,m}=\sum_{p=0}^{N-1} \sum_{n=0}^{M-1} b_{p,n} s^{p,n}_{q,m}\,, 
\label{eq:Precoding}
\ee
where $s^{p,n}_{q,m}$ denotes the two-dimensional precoding sequences and $d_{q,m}$ the results of the precoding operation, respectively. The time-frequency grid is defined by the discrete time index $m\in\{0,\ldots, M-1\}$ and the discrete frequency index $q\in\{0,\ldots, N-1\}$. The quadrature amplitude modulation alphabet is denoted by $\mathcal{A}$. 

Let $\Vec{B}\in\mathcal{A}^{N \times M}$ denote the symbol matrix with elements $b_{p,n}$. Using the notation introduced in Sec.~\ref{sec:intro}, we define the symbol vector $\Vec{b}=\vects(\Vec{B})=\vects\big((b_{p,n})\big)\in\mathcal{A}^{MN \times 1}$, and the precoded symbol vector $\Vec{d}=\vects\big((d_{q,m})\big)$. 
Matrix $\Vec{S}=[\Vec{s}_{0,0}, \ldots, \Vec{s}_{N-1,0}, \Vec{s}_{0,1}, \ldots, \Vec{s}_{N-1,M-1}]\oT\in\mathbb{C}^{MN\times MN}$
collects all vectorized 2D precoding sequences $\Vec{s}_{p,n}=\vects\big((s_{q,m}^{p,n})\big)$ 
column-wise. With these definitions, we can write \eqref{eq:Precoding} in vector matrix notation as
\be
\Vec{d}=\Vec{S}\Vec{b}\,.
\label{eq:PrecodingVec}
\ee

\subsection{Orthogonal Precoding for OFDM}
We use the signal model introduced in \cite{Farhang17, RezazadehReyhani17} and replace the DSFT precoding with \eqref{eq:PrecodingVec},
\begin{equation}
\Vec{y}=\diag(\Vec{g})\diag(\Vec{w}')\Vec{S}\Vec{b}+\Vec{n}\,.
\label{eq:OPOTFS}
\end{equation}
In \eqref{eq:OPOTFS}, the term $\diag(\Vec{g})$ represents the combined result of OFDM modulation, the doubly selective channel and OFDM demodulation\footnote{We assume that, due to a proper OFDM system design, the doubly selective channel does not cause inter-symbol interference, and the inter-carrier interference is small enough to be neglected. These are realistic assumptions for vehicular communication systems such as IEEE 802.11p \cite{Zemen12a, Mecklenbrauker11}, see the detailed discussion in \cite[Sec. II]{Zemen12}.}. The time-variant frequency response is denoted by $\Vec{g}=\vects\big((g_{q,m})\big)$, the received samples after OFDM demodulation by $\Vec{y}=\vects\big((y_{q,m})\big)$, and additive white complex symmetric Gaussian noise by $\Vec{n}=\vects\big((n_{q,m})\big)$. Elements $n_{q,m}\sim\mathcal{CN}(0,\sigma_n^2)$ have zero mean and variance $\sigma_n^2$. The TX window is denoted by $\Vec{w}'=\vects\big((w_{q,m}')\big)$. We assume no channel state information is available at the TX side and use a rectangular window $\Vec{w}'=\Vec{1}_{MN}$. 

Furthermore, we define the effective precoding matrix
\be
\Vec{\tilde{S}}=\diag(\Vec{g})\Vec{S}\,,
\label{eq:EffectiveSpreading}
\ee
combining the effect of linear precoding and the doubly-selective channel. Hence, the final signal model is given by
\be
\Vec{y}=\diag(\Vec{g})\Vec{S}\Vec{b}+\Vec{n}=\Vec{\tilde{S}}\Vec{b}+\Vec{n}\,.
\label{eq:EffSignalModel}
\ee
The orthonormality of the precoding sequences is lost, due to the multiplicative effect of the doubly-selective wireless propagation channel $\Vec{g}$.

\subsection{Precoding Sequences} 
\label{sec:PrecodingSeq}
We investigate three exemplary basis function sets for OP. Two sets contain constant modulus sequences, the basis functions of the DSFT and the Walsh-Hadamard transform (WHT), respectively. One basis function set contains general orthonormal sequences, namely 2D discrete prolate spheroidal (DPS) sequences.

\subsubsection{Discrete Symplectic Fourier Transform}
The DSFT precoding sequences \cite{Hadani17} are defined as
\be
{s_{q,m}^{p,n}}^\text{(DSFT)}=\frac{1}{\sqrt{MN}}\me^{\mj 2\pi(mn/M-qp/N)}\,.
\ee

\subsubsection{Walsh-Hadamard Transform}
We define the WHT precoding matrix recursively:
\be
{\Vec{S}_2}^\text{(WHT)}=\frac{1}{\sqrt{2}}\left[ 
\begin{array}{cc}
1 & 1 \\
1 & -1 \\
\end{array}
\right]
\ee
and
\be
{\Vec{S}_{2r}}^\text{(WHT)}=\frac{1}{\sqrt{2}}\left[ 
\begin{array}{cc}
\Vec{S}^\text{(WHT)}_r & \Vec{S}^\text{(WHT)}_r \\
\Vec{S}^\text{(WHT)}_r & -\Vec{S}^\text{(WHT)}_r \\
\end{array}
\right]\,.
\ee
The columns of $\Vec{S}^\text{(WHT)}$ are denoted by $\Vec{s}_{p,n}^\text{(WHT)}$. 

\subsubsection{2D Discrete Prolate Spheroidal (2DDPS) Sequences}
We define a general set of orthonormal basis functions as product of two DPS sequences:
\be
{s_{q,m}^{p,n}}^\text{(2DDPS)}= u_{n,m}(W_\text{t},I_\text{t})u_{p,q}(W_\text{f},I_\text{f}) 
\ee
with $W_\text{t}=[-\nnuD, \nnuD]$, $W_\text{f}= [0, \nthetaP]$, $I_\text{t}=\{0,\ldots,M-1\}$ and $I_\text{f}= \{0,\ldots, N-1\}$. DPS sequences $u_{i,\ell}(W,I)$ are the solution to the eigenvalue problem \cite{Zemen07a}
\be
\label{eq:DPS}
\sum_{\ell=0}^{M-1}C_{\ell-m}(W)u_{i,\ell}(W,I)= \lambda_i(W,I) u_{i,m}(W,I) 
\ee
for $m\in I$ with $C_k(W) =\int_W\me^{\mj 2\pi k \nu}\dif \nu=\frac{1}{\mj 2 \pi k}(\me^{\mj 2 \pi k \nu_{2}}-\me^{\mj 2 \pi k \nu_{1}})$
and $W=[\nu_1, \nu_2]$. 

\subsection{Precoding Complexity}
The precoding operation \eqref{eq:PrecodingVec} requires $M^2N^2$ complex multiplications for one frame in the general case. The OFDM modulation requires $M N \log N$ complex multiplications \cite{Cooley67}. 

The DSFT can be implemented using two consecutive fast Fourier transforms (FFTs), hence $N M \log M + MN \log N= MN \log {MN}$ complex multiplications are required for the OP of a data frame. For a rectangular window the N-point FFT of the DSFT and the N-point inverse FFT of the OFDM modulator can be combined \cite{Farhang17}. Hence, only $N M \log M$ complex multiplications are needed. The equivalent situation exists at the RX side, if a rectangular window is used.

The fast WHT \cite{Shanks69} requires $\frac{MN}{2}\log \frac{MN}{2}$ complex additions. For the 2DDPS sequences the precoding complexity is equal to the one in the general case.

\section{Iterative Detection}
\label{sec:Detection}
\begin{figure}
\def\svgwidth{\columnwidth}
\tiny{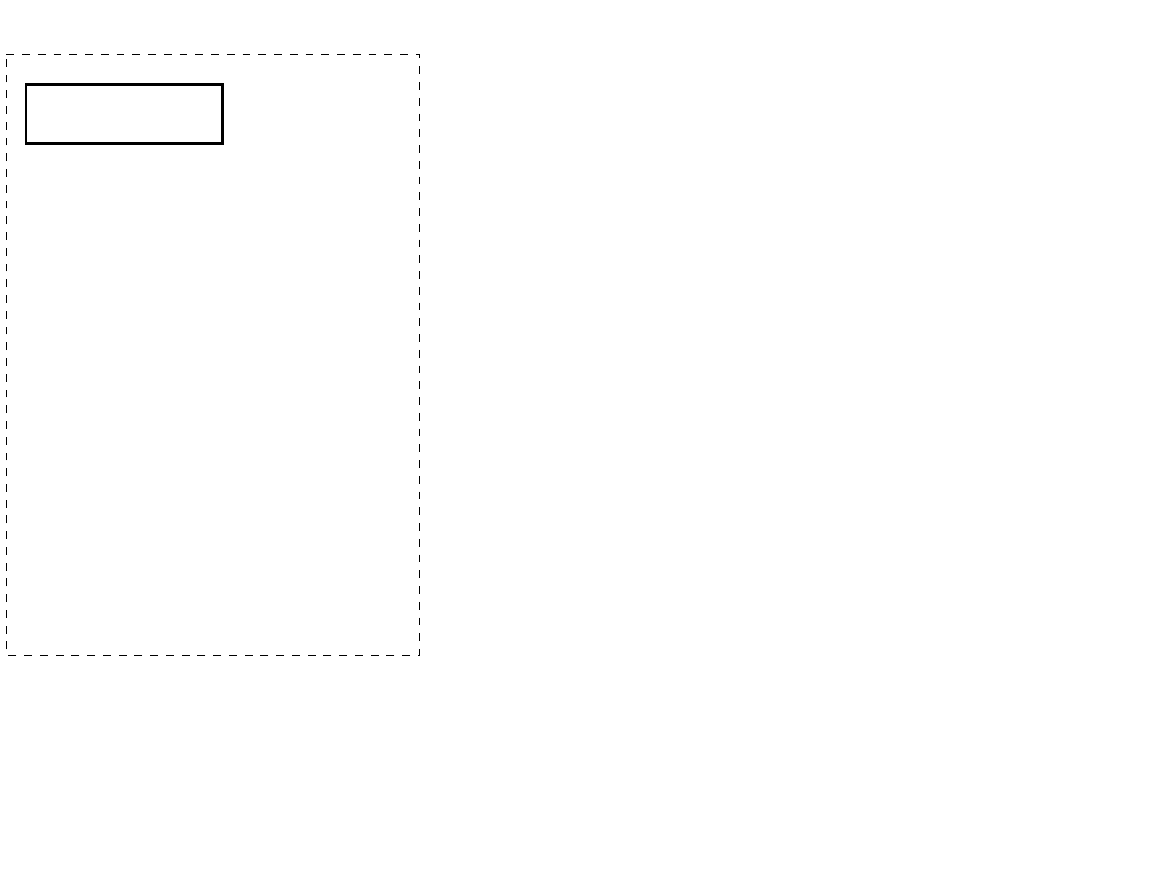}
\caption{System model of the OP TX and iterative OP RX.}
\label{fig:SystemModel}
\end{figure}

We use an iterative PIC algorithm \cite{Zemen06c} for symbol-wise maximum likelihood (ML) detection, using soft-symbol feedback. The a-posteriori probability (APP) output of the soft-input soft-output BCJR decoder \cite{BCJR74a} is interleaved and mapped to the alphabet $\mathcal{A}$ to obtain soft symbols $\tilde{b}_{p,n}$ \cite{Zemen06c}. The system model for the OP TX, the convolution with a doubly selective channel, and the OP RX is shown in Fig. \ref{fig:SystemModel}.

\subsection{Iteration $i=1$}
\label{sec:MMSE}
We obtain data symbol estimates 
\be
\hat{\Vec{b}}=\Vec{S}\oH\diag(\Vec{w}^{(1)})\Vec{y}\,,
\label{eq:ItOne}
\ee
by windowing the received samples $\Vec{y}$, followed by matched filtering $\Vec{S}\oH$. The vectorized receive window is denoted by $\Vec{w}^{(i)}=\vects\big((w^{(i)}_{q,m})\big)$ and the vectorized symbol estimates by $\Vec{\hat{b}}=\vects\big((\hat{b}_{q,m})\big)$, respectively. 

We perform windowing with a linear minimum mean-square error (LMMSE) filter
$w_{q,m}^{(1)}=\hat{g}_{q,m}^{*}/(|\hat{g}_{q,m}|^2+\sigma_n^2)$ in iteration $i=1$, to compensate for the lost orthogonality of $\Vec{\tilde{S}}$. Channel estimates, obtained through pilot information, are denoted by $\hat{g}_{q,m}$, see Section \ref{sec:ChannelEstimation}. 

Code-bit estimates $\hat{c}_k$, $k\in\{0,\ldots, MN|\mathcal\alpha|\}$, $\alpha=\log_2|\mathcal{A}|$, are obtained after demapping $\hat{b}_{p,n}$ and de-interleaving. We compute log-likelihood ratios (LLR) $L_k= 2\hat{c}_k/\sigma_n^2$ as input for the BCJR decoder. The BCJR decoder supplies estimates of the transmitted information bits, as well as APP for the code bits \cite{Zemen06c}.  
 
A partial diversity gain is obtained by OP detection with LMMSE equalization \cite{Mehana12} and matched filtering \eqref{eq:ItOne}. This is shown by numerical simulations later in Section \ref{sec:LinkLevelSim}. A low-complexity ML detection method is required for OP \cite{Hadani17} to fully utilize the diversity provided by the wireless propagation channel. Hence, we will use the soft-symbols from iteration $i-1$ for PIC in iteration $i$, to cancel the interference of other data symbols. This approach allows to implement a symbol-wise ML detector.

\subsection{Iteration $i>1$}
We express PIC for grid element $(p,n)$ according to
\begin{subequations}
\begin{align}
a_{p,n}^{(i)}&=\Vec{\tilde{s}}_{p,n}\oH\left(\diag(\Vec{w}^{(i)})\Vec{y}-\tilde{\Vec{S}}\tilde{\Vec{b}}^{(i-1)} + \tilde{\Vec{s}}_{p,n}\tilde{b}_{p,n}^{(i-1)}\right)\nonumber\\ \label{eq:PIC}
&\approx \underbrace{\Vec{\tilde{s}}_{p,n}\oH \tilde{\Vec{s}}_{p,n}}_{\gamma_{p,n}} b_{p,n} + \tilde{\Vec{s}}_{p,n}\oH \Vec{n}\\ 
&=\gamma_{p,n} b_{p,n} + \tilde{n}_{p,n}\,,
\label{eq:effSignalModel}
\end{align}
\end{subequations}
where \eqref{eq:PIC} becomes exact if we assume perfect PIC. The soft-symbol feedback vector $\tilde{\Vec{b}}^{(i)}=\vects\big((\tilde{b}_{p,n}^{(i)})\big)$ and the effective channel coefficient is denoted by $\gamma_{p,n}$. Noise $\tilde{n}_{p,n}$ has the same distribution as $n_{q,m}$. A rectangular receive window $\Vec{w}^{(i)}=\Vec{1}_{MN}$ is used for iterations $i>1$.

The symbol-wise ML expression 
\be 
\hat{b}_{p,n}=\argmin_{b_{p,n} \in\mathcal{A}}\{| a_{p,n} - \gamma_{p,n} b_{p,n} |^2\}
\label{eq:TFML}
\ee
for data symbol $b_{p,n}$ is formulated based on the scalar signal model \eqref{eq:effSignalModel}. A soft-output sphere decoder \cite{Studer08}, using \eqref{eq:TFML}, supplies LLRs $L_k$. The LLRs are used as input for the BCJR decoder. 

\section{Channel Hardening Through Precoding}
\label{sec:EffectiveChannel}
The performance of OP crucially depends on the distribution $f_\gamma(\gamma)$ of the effective channel coefficient $\gamma_{p,n}$ in \eqref{eq:effSignalModel}. We express $\gamma_{p,n}$ in element wise notation:
\begin{subequations}
\begin{align}
\gamma_{p,n} & = \Vec{\tilde{s}}_{p,n}\oH \tilde{\Vec{s}}_{p,n} = \\
&= \sum_{q=0}^{N-1}\sum_{m=0}^{M-1} {s^{p,n}_{q,m}}^* g_{q,m}^* g_{q,m} s^{p,n}_{q,m}=\\
&= \sum_{q=0}^{N-1}\sum_{m=0}^{M-1} |s^{p,n}_{q,m}|^2|g_{q,m}|^2\,.
\label{eq:gammaelement}
\end{align}
\end{subequations}

We treat two special cases:
\begin{enumerate}
\item No precoding (NO): In this case we can set $\Vec{S}=\Vec{I}$, i.e. $s_{q,m}^{p,n}=1$ for $p=q$ and $n=m$, otherwise $s_{q,m}^{p,n}=0$. We obtain $\gamma_{p,n}^\text{NO}= |g_{p,n}|^2$.
Here, the distribution of $\gamma_{p,n}^\text{NO}$ is determined by a single time-frequency grid-point, represented by $g_{p,n}$. Hence, $f_\gamma^\text{NO}$ will be exponentially distributed in Rayleigh fading channels, i.e., the diversity $d=1$ \cite{Molisch10}.

\item Precoding with constant modulus (CM) sequences: In this case $|s^{p,n}_{q,m}|^2=\frac{1}{MN}$
which applies, e.g., for the DSFT or WHT sequences. We obtain
\be
\gamma^\text{CM}_{p,n} = \gamma^\text{CM} = \sum_{q=0}^{N-1}\sum_{m=0}^{M-1} \frac{1}{MN} |g_{q,m}|^2\,.
\label{eq:gammaCM}
\ee
The distribution $f_\gamma^\text{CM}$ is determined by the average of all (correlated) channel samples $g_{q,m}$ on the time-frequency grid. All transmitted data symbols within a frame are effected by the same effective channel coefficient $\gamma^\text{CM}$. See also the treatment of maximum ratio combining of correlated diversity sources in \cite{Molisch10}. Due to the averaging in \eqref{eq:gammaCM}, the fading effect is strongly reduced, resulting in channel hardening in the time-frequency domain. 
\end{enumerate}

OP for doubly-selective channels does \emph{not} require the usage of the DSFT, as proven by \eqref{eq:gammaCM}. Hence, a transform between the time-frequency and the delay-Doppler domain is not required for the precoding of data symbols in doubly selective channels. \emph{All} constant modulus sequences will provide the same performance. 

The conclusion above does not apply for channel estimation. The transform to the delay-Doppler domain might be beneficial for channel estimation, if precoded pilot-symbols shall be used and the doubly selective channel exhibits a sparse delay-Doppler representation \cite{Li17}.

\subsection{Quantifying Channel Hardening by Orthogonal Precoding}
\label{sec:ChannelHardening}
We describe the distribution $f_\gamma^\text{CM}$ for a given doubly-selective fading process and constant modulus OP sequences. The channel hardening effect of OP is quantified by means of the second moment of $f_\gamma^\text{CM}$.

We express $\Vec{g}$ as filtered white random process,
\be
\Vec{g}=\Vec{U}\sqrt{\Vec{\Sigma}} \Vec{U}\oH\Vec{z}\,,
\ee
where $\Vec{z}\sim \mathcal{CN}(0,\Vec{I}_{MN})$ is a complex Gaussian random vector with independent identically distributed (i.i.d.) entries. 
Matrix $\Vec{U}$ contain eigenvectors of the covariance matrix $\Vec{R}_g$, and $\Vec{\Sigma}$ contains eigenvalues $\lambda_i$ on the main diagonal:
\be
\Vec{R}_g=\E\{\Vec{g}\Vec{g}\oH\}=\Vec{U}\Vec{\Sigma} \Vec{U}\oH\,.
\label{eq:rgeig}
\ee

With these prerequisites we write \eqref{eq:gammaCM} as
\begin{subequations}
\begin{align}
\gamma^\text{CM} &= \frac{1}{MN}\Vec{g}\oH\Vec{g} =\\
&=\frac{1}{MN}\Vec{z}\oH \Vec{U}\sqrt{\Vec{\Sigma}} \Vec{U}\oH\Vec{U}\sqrt{\Vec{\Sigma}} \Vec{U}\oH \Vec{z} = \\
&= \frac{1}{MN}\Vec{z}\oH\Vec{U}\Vec{\Sigma} \Vec{U}\oH \Vec{z} = \frac{1}{MN}\Vec{\tilde{z}}\oH\Vec{\Sigma}\Vec{\tilde{z}}= \\
&= \frac{1}{MN}\sum_{i=0}^{MN-1} \lambda_i |\tilde{z}_i|^2\,,
\label{eq:CMeffch}
\end{align}
\end{subequations}
where $\Vec{\tilde{z}}\sim\mathcal{CN}(0,\Vec{I}_{MN})$, since $\Vec{U}$ is an unitary matrix. Hence, $\gamma^\text{CM}$ is distributed according to a sum of exponentially distributed random variables weighted by eigenvalues $\lambda_i$. A closed form expression for $f_\gamma^\text{CM}$ is given by \cite{Akkouchi08}. The variance
\be
{\sigma_\gamma^2}=\frac{1}{M^2N^2}\sum_{i=0}^{MN -1}\lambda_i^2
\label{eq:gammaVariance} 
\ee
of $\gamma^\text{CM}$ depends on the eigenvalue distribution of $\Vec{R}_g$, and is a direct measure of the channel hardening effect of OP.

The direct evaluation of \eqref{eq:rgeig} for a general doubly-selective correlated fading processes is numerically difficult, due to the multiplicity of the largest eigenvalues of $\Vec{R}_g$. A numerically stable algorithm for the calculation of $\lambda_i$ for a fading process with a flat delay-Doppler scattering function is shown in \cite{Zemen12a}. This fading process is defined by two parameters, the normalized delay support $\nthetaP$ and the normalized Doppler support $\nnuD$. The correlation matrix for the flat delay-Doppler scattering function is denoted by $\Vec{\tilde{R}}_g(\nnuD, \nthetaP)$, see \cite[(26)-(28)]{Zemen12a}. We use the algorithm from \cite{Zemen12a} to calculate the eigenvalues $\tilde{\lambda}_i$ of $\Vec{\tilde{R}}_g(\nnuD, \nthetaP)$.

\subsection{Numerical Examples}
\label{sec:NumExp}
We evaluate the distribution of $\tilde{\lambda}_i$ for the OFDM parameters of the IEEE 802.11p OFDM physical layer: $N=64$ subcarriers, $M=44$ OFDM symbols frame length, bandwidth $B=10\,\text{MHz}$, carrier frequency $\nfC=5.9\,\text{GHz}$, and cyclic prefix length $G=16$. 

We define four channel types in Table~\ref{tab:Channel}, using $\nnuD$ and $\nthetaP$ to describe the normalized support of the Doppler-spectral density (DSD) and the power-delay profile (PDP). The maximum relative velocity between TX and RX is $200\,\text{km/h}$, i.e. the maximum Doppler shift is $1092\,\text{Hz}$. The maximum support of the PDP is set to $1\,\mu\text{s}$. Both values are characteristic for urban vehicular environments \cite{Bernado14}.
\begin{table}
\begin{center}
\caption{Normalized delay and Doppler support for four channel types with corresponding variance of the effective channel coefficient $\gamma$.}
\label{tab:Channel}
\begin{tabular}{llll} 
\toprule
Type	      				& $\nnuD$    & $\nthetaP$  		&    $\sigma_\gamma^2$ \\
\midrule
\emph{non-selective}		& 	 $0.0001$ & $0.0001$		&    1    \\
\emph{time-selective}		& 	 $0.009$	  & $0.0001$		&	 0.76 \\
\emph{frequency-selective}  &    $0.0001$ & $0.15$			&	 0.09  \\
\emph{doubly-selective} 	& 	 $0.009$	  & $0.15$			& 	 0.07 \\
\bottomrule
\end{tabular}
\end{center}
\end{table}

\begin{figure}
	\centering
	\includegraphics[width=\columnwidth]{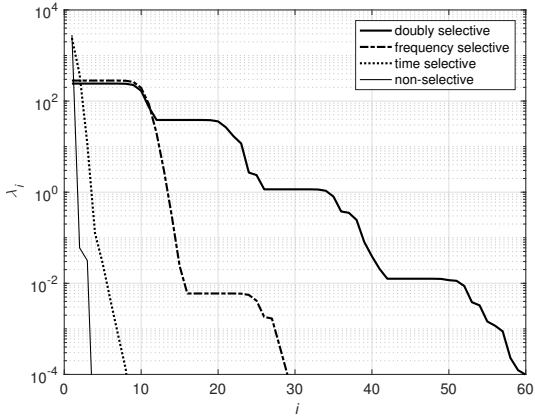}
	\caption{Eigenvalue distribution of $\Vec{\tilde{R}}_g$ for four channel types defined in Table~\ref{tab:Channel}.} 
	\label{fig:EigenvalueDistr}
\end{figure}

The eigenvalue distribution of $\Vec{\tilde{R}}_g(\nnuD, \nthetaP)$, for the four channel types of Tab.~\ref{tab:Channel}, is depicted in Fig.~\ref{fig:EigenvalueDistr}. 
Matrix $\Vec{\tilde{R}}_g(\nnuD, \nthetaP)$ is full rank and has $MN$ eigenvalues. We plot all eigenvalues $\tilde{\lambda}_i>10^{-4}$.
We use \eqref{eq:gammaVariance} to compute ${\sigma_\gamma^2}$ in the fourth column of Table~\ref{tab:Channel}. The channel hardening effect of OP increases with increasing support of the PDP and the DSD, i.e., the variance ${\sigma_\gamma^2}$ decreases. For a doubly selective channel the eigenvalues decay the slowest, see Fig.~\ref{fig:EigenvalueDistr}, and the variance $\sigma_\gamma^2$ is the smallest, see Tab.~\ref{tab:Channel}.

\section{Iterative Channel Estimation}
\label{sec:ChannelEstimation}
Channel estimates are required for coherent detection at the RX side. Hence, we interleave $S_\text{p}$ pilot symbols $\Vec{p}\in\mathbb{C}^{S_\text{p} \times 1}$ with $S_\text{d}$ precoded data symbols $\Vec{d}\in\mathbb{C}^{S_\text{d} \times 1}$ in the time-frequency domain, such that $S_\text{d}+S_\text{p}= MN$. We modify the signal model \eqref{eq:EffSignalModel} and express the interleaving with a permutation matrix $\Vec{P}$:
\be
\Vec{y}=\diag(\Vec{g})\underbrace{[\Vec{P}_\text{p} \Vec{P}_\text{d}]}_\Vec{P} \left[\begin{array}{c}
\Vec{p} \\
\Vec{S}\Vec{b}
\end{array}
\right]
+\Vec{n}\,,
\label{eq:GOTFSFinal}
\ee
where $\Vec{P}_\text{p}\in\mathbb{R}^{MN \times S_\text{p}}$ describes the pilot symbol placement and $\Vec{P}_\text{d}\in\mathbb{R}^{MN \times S_\text{d}}$ the (precoded) data symbol positions in the time-frequency grid.  All the equations in the previous Sections \ref{se:SignalModel}-\ref{sec:EffectiveChannel} are still valid by replacing $M$ with $M'$, $N$ with $N'$, and $\Vec{g}$ with $\Vec{P}_\text{d}\Vec{g}$, where $M'\le M$, $N' \le N$, and $S_\text{d}=M'N'$.

For channel estimation, we rewrite \eqref{eq:GOTFSFinal} as
\be
\Vec{y}=\diag\left(\Vec{P}_\text{p} \Vec{p} + \Vec{P}_\text{d} \Vec{S}\Vec{b}\right)\Vec{g}+\Vec{n}\,.
\label{eq:GOTFSChannelEst}
\ee
Following the derivation in \cite[(30)-(39)]{Zemen06c}, we obtain the Wiener filter
for $\Vec{g}$ as
\be
\label{eq:WienerFilter}
\hat{\Vec{g}}=\Vec{R}_g{\Vec{\tilde{D}}}\oH\left(\Vec{\tilde{D}}\Vec{R}_g{\Vec{\tilde{D}}}\oH+\Vec{\Lambda}+\sigma^2_z\Vec{I}_{MN}\right)^{-1}\Vec{y}\,.
\ee
The orthogonal precoded soft-symbol feedback is expressed as
\be
\Vec{\tilde{D}}=\diag\big(\Vec{P}_\text{p} \Vec{p} + \Vec{P}_\text{d} \Vec{S}\Vec{\tilde{b}}\big)\,,
\ee 
and 
\be
\Vec{\Lambda}=\diag(\Vec{P}_\text{p} \Vec{0}_{S_\text{p}}+ \Vec{P}_\text{d} \Vec{1}_{S_\text{d}}(1-\sigma^2_{\tilde{b}}))\,. 
\ee
For time-frequency grid positions $(q,m)$ where pilot symbols are transmitted, the corresponding entries on the diagonal of $\Vec{\Lambda}$ are equal to zero. All other entries, related to precoded data-symbols, are filled with the variance of the soft-symbol feedback of the BCJR decoder. 

The soft-symbol feedback is modeled as complex Gaussian distributed $\Vec{\tilde{b}}\sim \mathcal{CN}(0,\sigma^2_{\tilde{b}}\Vec{I}_{S_\text{d}})$, with zero mean and variance $\sigma^2_{\tilde{b}}\Vec{I}_{S_\text{d}}$. Hence, if the output of the BCJR decoder converges towards the true transmit symbols, the term $(1-\sigma^2_{\tilde{b}})$ tends to zero, and \eqref{eq:WienerFilter} becomes a classic Wiener filter. The sample variance of the soft-symbol feedback is estimated according to $\hat{\sigma}^2_{\tilde{b}}=\frac{1}{S_\text{d}}\sum_{p=0}^{N'-1}\sum_{n=0}^{M'-1}|\tilde{b}_{p,n}|^2$. The eigenvalue structure of $\Vec{R}_g$ can be exploited to implement a reduced rank version of \eqref{eq:WienerFilter}, reducing the numerical complexity, see \cite[(32)]{Zemen12a}. 

\section{Numerical Link Level Simulation} 
\label{sec:LinkLevelSim}
\subsection{Simulation Parameters}
We use the same parameters as in Sec. \ref{sec:NumExp}. The pilot pattern is adapted for vehicular velocities \cite{Mecklenbrauker11} by using $J=4$ dedicated pilot OFDM symbols distributed over the OFDM frame\footnote{The IEEE 802.11p pilot pattern does not allow aliasing free channel estimation in non line-of-sight scenarios \cite{Mecklenbrauker11}. An iterative channel estimation algorithm is required for velocities higher than $50\,\text{km/h}$ \cite{Zemen12, Zemen12a}.} at time-indices
$m\in\left\{\left\lfloor i\frac{M}{J} + \frac{M}{2J}\right\rfloor|\, i\in\{0,\ldots, J-1\} \right\}=\{5, 16, 27, 38\}$. A quadrature phase shift keying (QPSK) symbol alphabet $\mathcal{A}$ is used and a $r=1/2$ convolutional code for channel coding followed by a random interleaver.

\subsection{Geometry Based Channel Model}
\label{sec:GSCM}
We use a geometry-based channel model (GCM) for the numerical simulation of a doubly selective communication channel. The non-stationary fading process \cite{Bernado14} is approximated as wide-sense stationary for the duration of $M$ OFDM symbols, $m\in\{0,\ldots, M-1\}$, and $N$ subcarriers, $q\in\{0,\ldots, N-1\}$. Hence, we model the time-variant path delay as $\tau_\ell(t)=\tau_\ell(0) - f_\ell t/\nfC$ for the duration of $M \nTS$, where $f_\ell$ denotes the Doppler shift of path $\ell$, $\nfC$ the carrier frequency, and $\nTS=\nTC (N+G)$ denotes the duration of an OFDM symbol. The length of the cyclic prefix in samples is denoted by $G$, the chip duration $\nTC=1/B$, and $B$ denotes the system bandwidth.

The sampled time-variant frequency-response $g_{q,m}$ is defined as 
\be
g_{q,m}=g^\text{(TX)}_qg^\text{(RX)}_q\sum_{\ell=0}^{P-1}\eta_\ell{\me^{-\mj2\pi \theta_\ell q}} \me^{\mj2\pi \nu_\ell m }\,,
\label{eq:GSCM}
\ee
where $\nu_\ell=f_\ell \nTS$ denotes the normalized Doppler shift, $\theta_\ell=\tau_\ell(0)/(N\nTC)$ the normalized path delay, $\eta_\ell$ the path weight, and $P$ the number of propagation paths, respectively. The Doppler shift is bounded by the maximum normalized Doppler support $\nnuD$, i.e., $|\nu_\ell|\le \nnuD$. The normalized path delay is bounded by the maximum normalized delay support $\nthetaP$, i.e., $0\le \theta_\ell\leq\nthetaP$. The TX- and RX-filter is denoted by $g^\text{(TX)}_q$ and $g^\text{(RX)}_q$, respectively.

The GCM is parametrized such that the fading process has an exponentially decaying PDP with a root-mean square delay spread of $0.4\,\mu\text{s}$, and a flat DSD according to a relative velocity of $v\in\{0,200\}\,\text{km/h}$. These parameters are a realistic assumption in many traffic scenarios, for vehicle-to-vehicle or vehicle-to-infrastructure links \cite{Mecklenbrauker11}.

\subsection{Link Level Simulation Results}
\begin{figure} 
	\centering
	\includegraphics[width=\columnwidth]{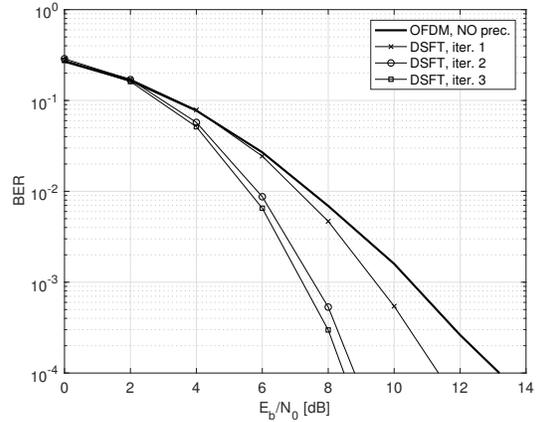}
	\caption{BER versus $\nEb/N_0$ comparing OFDM with DSFT OP for perfect CSI and velocity of $v=200\,\text{km/h}$. The iterative OP RX uses three iterations.}
	\label{fig:OPDSFTPerfCSI}
\end{figure}

We show the bit error rate (BER) versus $\nEb/N_0$ for an OFDM physical layer without precoding in Fig.~\ref{fig:OPDSFTPerfCSI}. This result is compared with DSFT OP using three iterations for PIC. Perfect channel state information (CSI) is available at the RX-side. In the first iteration, our OP RX uses LMMSE equalization, which lacks full diversity. The results of the second iteration are obtained with soft-symbol PIC and symbol-wise ML decoding with a soft-output sphere decoder.  From iteration one to iteration two, a gain of about $2.5\,\text{dB}$ for a BER of $10^{-4}$ and a velocity of $v=200\,\text{km/h}$ is presented in Figure~\ref{fig:OPDSFTPerfCSI}. An additional gain of about $0.3\,\text{dB}$ is obtained with iteration three. No performance improvement can be achieved for more than three iterations. The total gain of OP compared to OFDM is $4.8\,\text{dB}$.

\begin{figure}
	\centering
	\includegraphics[width=\columnwidth]{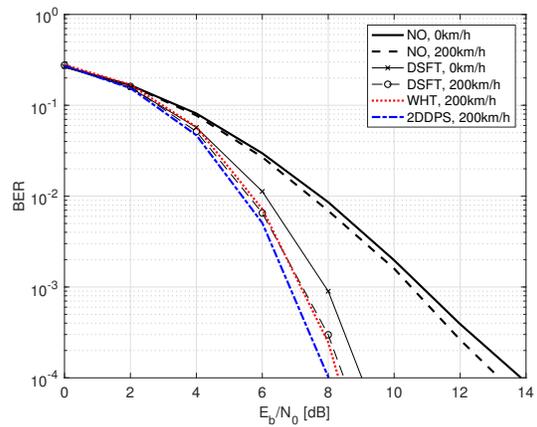}
	\caption{BER versus $\nEb/N_0$ comparing no precoding (NO) with OP using DSFT, 2DDPS and WHT sequences for perfect CSI and velocities $v\in\{0, 200\}\,\text{km/h}$. All OP results are shown after three decoder iterations.}
	\label{fig:OPcomparison}
\end{figure}

We provide simulation results for OFDM as well as for DSFT OP after the third iteration with $v\in\{0, 200\}\,\text{km/h}$ in Fig.~\ref{fig:OPcomparison}. Furthermore, we show results for OP with the WHT and 2DDPS sequences. Both constant modulus sequences, the DSFT and the WHT sequences, result in the same performance, as predicted by \eqref{eq:gammaCM}. The 2DDPS sequences show the best performance, being about $0.2\,\text{dB}$ better for a BER of $10^{-4}$, compared to the DSFT and WHT sequences.

\begin{figure}
	\centering
	\includegraphics[width=\columnwidth]{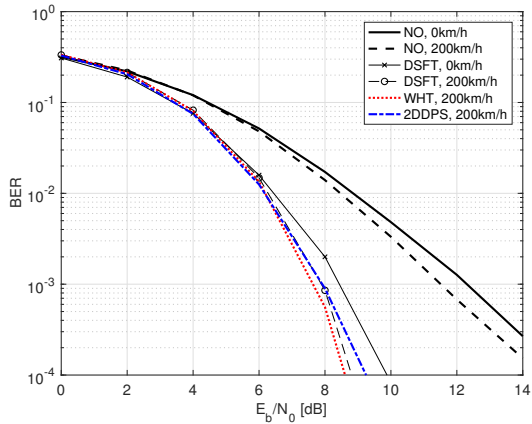}
	\caption{BER versus $\nEb/N_0$ comparing no precoding (NO) with OP using DSFT, 2DDPS and WHT sequences. CSI is estimated using pilot symbols interleaved with data symbols, the velocity $v\in\{0, 200\}\,\text{km/h}$. All OP results are shown after three decoder iterations.}
	\label{fig:OPcomparisonCHest}
\end{figure}

Finally, in Fig.~\ref{fig:OPcomparisonCHest}, we show OP results using channel estimates. All results are shifted to the right by $1\,\text{dB}$, due to the error in the CSI. This CSI error causes also non perfect PIC, leading to slight changes in the OP results.

\section{Conclusions} 
\label{sec:Conclusions}
We presented a signal model for general OP using OFDM as time-frequency modulation method. Iterative PIC and iterative channel estimation methods are applied for OP. For the first time in the open literature, a RX for OP using \emph{coded transmission} is presented in detail. We proved that all constant modulus sequences, e.g. DSFT or WHT sequences, lead to the same performance for OP. We demonstrated that the channel hardening effect of OP increases with larger delay- and Doppler spread of the channel impulse response. We compared the performance of DSFT, WHT, and 2DDPS OP sequences, using a physical layer close to the one of IEEE 802.11p in vehicular scenarios. We demonstrate, that our OP RX achieves a gain of about $4.8\,\text{dB}$ for a bit error rate of $10^{-4}$ and a relative velocity of $0\ldots200\,\text{km/h}$ if compared to OFDM. The presented performance result for coded OP are the best results published so far for a fully documented RX architecture, allowing to verfiy and extend our research results.

\input{OPPIMRCfinal.bbl}

\end{document}

%% file: OPV2.pdf_tex
\begingroup%
  \makeatletter%
  \providecommand\color[2][]{%
    \errmessage{(Inkscape) Color is used for the text in Inkscape, but the package 'color.sty' is not loaded}%
    \renewcommand\color[2][]{}%
  }%
  \providecommand\transparent[1]{%
    \errmessage{(Inkscape) Transparency is used (non-zero) for the text in Inkscape, but the package 'transparent.sty' is not loaded}%
    \renewcommand\transparent[1]{}%
  }%
  \providecommand\rotatebox[2]{#2}%
  \ifx\svgwidth\undefined%
    \setlength{\unitlength}{336.45987841bp}%
    \ifx\svgscale\undefined%
      \relax%
    \else%
      \setlength{\unitlength}{\unitlength * \real{\svgscale}}%
    \fi%
  \else%
    \setlength{\unitlength}{\svgwidth}%
  \fi%
  \global\let\svgwidth\undefined%
  \global\let\svgscale\undefined%
  \makeatother%
  \begin{picture}(1,0.74651651)%
    \put(0,0){\includegraphics[width=\unitlength,page=1]{OPV2.pdf}}%
    \put(0.10538264,0.64027556){\color[rgb]{0,0,0}\makebox(0,0)[b]{\smash{Encoder}}}%
    \put(0,0){\includegraphics[width=\unitlength,page=2]{OPV2.pdf}}%
    \put(0.49069309,0.37897901){\color[rgb]{0,0,0}\makebox(0,0)[lb]{\smash{$i$=1: LMMSE+matched filter}}}%
    \put(0,0){\includegraphics[width=\unitlength,page=3]{OPV2.pdf}}%
    \put(0.57717785,0.21902985){\color[rgb]{0,0,0}\makebox(0,0)[b]{\smash{OFDM demodulation }}}%
    \put(0.73816839,0.65674479){\color[rgb]{0,0,0}\makebox(0,0)[b]{\smash{APP}}}%
    \put(0.59592404,0.28075701){\color[rgb]{0,0,0}\makebox(0,0)[lb]{\smash{$y_{q,m}$}}}%
    \put(0,0){\includegraphics[width=\unitlength,page=4]{OPV2.pdf}}%
    \put(0.25946579,0.70945932){\color[rgb]{0,0,0}\makebox(0,0)[b]{\smash{Transmitter}}}%
    \put(0,0){\includegraphics[width=\unitlength,page=5]{OPV2.pdf}}%
    \put(0.10538264,0.55602642){\color[rgb]{0,0,0}\makebox(0,0)[b]{\smash{Interleaver}}}%
    \put(0,0){\includegraphics[width=\unitlength,page=6]{OPV2.pdf}}%
    \put(0.11429182,0.43164886){\color[rgb]{0,0,0}\makebox(0,0)[lb]{\smash{$b_{p,n}$}}}%
    \put(0,0){\includegraphics[width=\unitlength,page=7]{OPV2.pdf}}%
    \put(0.10505324,0.47235911){\color[rgb]{0,0,0}\makebox(0,0)[b]{\smash{Symbol mapper}}}%
    \put(0,0){\includegraphics[width=\unitlength,page=8]{OPV2.pdf}}%
    \put(0.10505322,0.38810997){\color[rgb]{0,0,0}\makebox(0,0)[b]{\smash{Orth. precoding}}}%
    \put(0,0){\includegraphics[width=\unitlength,page=9]{OPV2.pdf}}%
    \put(0.10538264,0.30327899){\color[rgb]{0,0,0}\makebox(0,0)[b]{\smash{Pilot insertion}}}%
    \put(0,0){\includegraphics[width=\unitlength,page=10]{OPV2.pdf}}%
    \put(0.11429182,0.34732275){\color[rgb]{0,0,0}\makebox(0,0)[lb]{\smash{$d_{q,m}$}}}%
    \put(0,0){\includegraphics[width=\unitlength,page=11]{OPV2.pdf}}%
    \put(0.22839127,0.30849513){\color[rgb]{0,0,0}\makebox(0,0)[lb]{\smash{$\Vec{p}$}}}%
    \put(0,0){\includegraphics[width=\unitlength,page=12]{OPV2.pdf}}%
    \put(0.10538264,0.21902985){\color[rgb]{0,0,0}\makebox(0,0)[b]{\smash{OFDM modulation}}}%
    \put(0,0){\includegraphics[width=\unitlength,page=13]{OPV2.pdf}}%
    \put(0.52889271,0.12725355){\color[rgb]{0,0,0}\makebox(0,0)[b]{\smash{+}}}%
    \put(0.32274196,0.08784502){\color[rgb]{0,0,0}\makebox(0,0)[lb]{\smash{$h_{q,m}$}}}%
    \put(0.53977918,0.08920249){\color[rgb]{0,0,0}\makebox(0,0)[lb]{\smash{$n_{q,m}$}}}%
    \put(0,0){\includegraphics[width=\unitlength,page=14]{OPV2.pdf}}%
    \put(0.3088598,0.12760433){\color[rgb]{0,0,0}\makebox(0,0)[b]{\smash{Convolution}}}%
    \put(0.30835914,0.0434813){\color[rgb]{0,0,0}\makebox(0,0)[b]{\smash{Impuls resp.}}}%
    \put(0,0){\includegraphics[width=\unitlength,page=15]{OPV2.pdf}}%
    \put(0.52550866,0.0429237){\color[rgb]{0,0,0}\makebox(0,0)[b]{\smash{Noise}}}%
    \put(0,0){\includegraphics[width=\unitlength,page=16]{OPV2.pdf}}%
    \put(0.57717785,0.64027556){\color[rgb]{0,0,0}\makebox(0,0)[b]{\smash{BCJR decoder}}}%
    \put(0,0){\includegraphics[width=\unitlength,page=17]{OPV2.pdf}}%
    \put(0.57717785,0.55602642){\color[rgb]{0,0,0}\makebox(0,0)[b]{\smash{Deinterleaver}}}%
    \put(0,0){\includegraphics[width=\unitlength,page=18]{OPV2.pdf}}%
    \put(0.59574162,0.43266926){\color[rgb]{0,0,0}\makebox(0,0)[lb]{\smash{$\hat{b}_{p,n}$}}}%
    \put(0,0){\includegraphics[width=\unitlength,page=19]{OPV2.pdf}}%
    \put(0.57717785,0.47235911){\color[rgb]{0,0,0}\makebox(0,0)[b]{\smash{Symbol demapper}}}%
    \put(0,0){\includegraphics[width=\unitlength,page=20]{OPV2.pdf}}%
    \put(0.49069309,0.34564566){\color[rgb]{0,0,0}\makebox(0,0)[lb]{\smash{$i$$>$1: PIC+soft sphere decoder}}}%
    \put(0,0){\includegraphics[width=\unitlength,page=21]{OPV2.pdf}}%
    \put(0.90060516,0.3212591){\color[rgb]{0,0,0}\makebox(0,0)[b]{\smash{Channel est.}}}%
    \put(0,0){\includegraphics[width=\unitlength,page=22]{OPV2.pdf}}%
    \put(0.89785115,0.55602642){\color[rgb]{0,0,0}\makebox(0,0)[b]{\smash{Interleaver}}}%
    \put(0,0){\includegraphics[width=\unitlength,page=23]{OPV2.pdf}}%
    \put(0.77138353,0.40057273){\color[rgb]{0,0,0}\makebox(0,0)[lb]{\smash{$\tilde{b}_{p,n}$}}}%
    \put(0,0){\includegraphics[width=\unitlength,page=24]{OPV2.pdf}}%
    \put(0.8975218,0.47235911){\color[rgb]{0,0,0}\makebox(0,0)[b]{\smash{Symbol mapper}}}%
    \put(0,0){\includegraphics[width=\unitlength,page=25]{OPV2.pdf}}%
    \put(0.77132651,0.29168162){\color[rgb]{0,0,0}\makebox(0,0)[lb]{\smash{$\hat{h}_{q,m}$}}}%
    \put(0.91170195,0.70975869){\color[rgb]{0,0,0}\makebox(0,0)[b]{\smash{Receiver<}}}%
    \put(0.1075509,0.72845157){\color[rgb]{0,0,0}\makebox(0,0)[b]{\smash{bits}}}%
    \put(0.57661838,0.72700607){\color[rgb]{0,0,0}\makebox(0,0)[b]{\smash{bits}}}%
    \put(0.12142025,0.60044718){\color[rgb]{0,0,0}\makebox(0,0)[lb]{\smash{$c[k]$}}}%
    \put(0.59396919,0.60060255){\color[rgb]{0,0,0}\makebox(0,0)[lb]{\smash{$L_k$}}}%
  \end{picture}%
\endgroup%